\documentstyle[12pt]{article}

\begin{document}

\LaTeX{}

\begin{center}
The equation of state of a degenerate Fermi gas\bigskip\ 

Vladan Celebonovic\medskip\ 

Institute of Physics,Pregrevica 118,11080 Zemun-Beograd,Yugoslavia\smallskip%
\ 

celebonovic@exp.phy.bg.ac.yu

vcelebonovic@sezampro.yu\bigskip\ 
\end{center}

Abstract: An analytical expression for Fermi-Dirac integrals of arbitrary
order is presented,and its applicability in obtaining an analytical EOS of a
degenerate non-relativistic Fermi gas is discussed to some extent.

\begin{center}
Introduction
\end{center}

Degenerate Fermi gases occur in a variety of systems studied in astrophysics
and ''laboratory'' physics.Examples of such systems include white dwarfs and
neutron stars,planetary interiors,ordinary metals and organic conductors.

The general form of the equation of state (EOS) of a non-relativistic
degenerate Fermi gas has been determined more than half a century ago \qquad
( for example,Chandrasekhar,1939) as:

\begin{equation}
\label{(1)}n_e=\frac{4\pi }{h^3}(2m_ek_BT)^{3/2}F_{1/2}(\eta ) 
\end{equation}
\bigskip\ 

All the symbols have their standardized meanings ,and $F_{1/2}(\eta )$
denotes a particular case of the Fermi-Dirac ( FD\ ) integrals of order n:

\begin{equation}
\label{(2)}F_n(\eta )=\int_0^\infty \frac{f(\epsilon )d\epsilon }{1+\exp
[\beta (\epsilon -\mu )]} 
\end{equation}
\medskip\ 

where $f(\epsilon )=\epsilon ^n$,n$\in $R,$\mu $ denotes the chemical
potential,$\beta $ is the inverse temperature,and $\eta =\beta \mu .$

\newpage\ 

The practical problem of the numerical evaluation of the FD integrals has
been the object of several recent studies ( such as Cloutman,1989;
Antia,1993; Miralles and Van Riper,1995).The purpose of the present paper is
to obtain an analytical approximation to the FD\ integrals of arbitrary
order,and to apply it to the problem of the EOS of a degenerate Fermi gas.%
\bigskip\ 

\begin{center}
The calculations\smallskip\ 
\end{center}

Taking $k_B=1,$and introducing a change of variables by

\begin{equation}
\label{(3)}\epsilon -\mu =Tz 
\end{equation}

eq.(2) can be transformed into the following form: 
\begin{equation}
\label{(4)}F_n(\beta \mu )=T\int_{-\mu /T}^\infty \frac{f(\mu +Tz)}{1+\exp
[z]}=\int_0^\mu f(\epsilon )d\epsilon +T\int_0^\infty \frac{f(\mu +Tz)-f(\mu
-Tz)}{1+\exp [z]}dz 
\end{equation}
\medskip\ The difference of values of $f$ in the last integral of eq.(4) can
be developed into series as

\begin{equation}
\label{(5)}f(\mu +Tz)-f(\mu -Tz)=\sum_{n=0}^\infty [1-(-1)^n]\frac{(Tz)^n}{n!%
}f^{(n)}(\mu ) 
\end{equation}
\medskip\ Inserting eq.( 5 ) into eq.(4 ) and using the fact that%
$$
\int_0^\infty \frac{x^{\alpha -1}}{1+\exp [z]}dx=(1-2^{1-\alpha })\Gamma
(\alpha )\zeta (\alpha ) 
$$
\medskip\ where $\Gamma (\alpha )$ and $\zeta (\alpha )$ denote the gamma
function and Riemann's zeta function,one gets the following final form of
the FD integrals of arbitrary order: 
\begin{equation}
\label{(6)}F_n(\beta \mu )=\int_0^\mu f(\epsilon )d\epsilon
+T\sum_{n=0}^\infty \frac{f^{(n)}(\mu )}{n!}[1-(-1)^n]T^n(1-2^{-n})\Gamma
(n+1)\zeta (n+1) 
\end{equation}
\medskip\ This expression is a generalization to arbitrary order and number
of terms of existing partial results ( for example,Landau and Lifchitz,1976
).\newpage\ 

\begin{center}
Discussion\medskip\ 
\end{center}

Inserting $n=1/2$ into eq.( 6 ) and limiting the sum to terms up to and
including $T^6$ ,it follows that

\begin{equation}
\label{(7)}F_{1/2}(\eta )\cong \frac 23\mu ^{3/2}[1+\frac{\pi ^2}8(\frac
T\mu )^2+\frac{7\pi ^4}{640}(\frac T\mu )^4+\frac{31\pi }{3072}(\frac T\mu
)^2] 
\end{equation}
\medskip\ This result is impractical,because the chemical potential itself
is a function of T and the number density of the Fermi gas.It can be shown
from eq.(1) that this function can be approximated by the following
developement: 
\begin{equation}
\label{(8)}\mu \cong \mu _0[1-\frac{\pi ^2}{12}(\frac T{\mu _0})^2+\frac{\pi
^4}{720}(\frac T{\mu _0})^4-\frac{\pi ^6}{162}(\frac T{\mu _0})^6+\frac{\pi
^8}{754}(\frac T{\mu _0})^8] 
\end{equation}
\medskip\ The symbol $\mu _0$ denotes the chemical potential of the electron
gas at T = 0 K,and it depends only on the particle number density.Inserting
eq.( 8 ) into eq.(7) and developing into series in T,one gets 
\begin{equation}
\label{(9)}F_{1/2}(T)\cong \frac 23\mu _0^{3/2}[1+\frac{\pi ^4}{48}(\frac
T{\mu _0})^4+\frac{19\pi ^6}{5760}(\frac T{\mu _0})^6+\frac{\pi ^8}{146}%
(\frac T{\mu _0})^8] 
\end{equation}
\medskip\ Note that we have managed to express this FD integral as an
explicite function of the particle number density and the temperature.This
result is a distnct advantage over some previous numerical work ( such as
Cloutman,1989 ),where the argument of this integral was left in the form $%
\beta \mu ,$without taking into account the number-density and temperature
dependences of the chemical potential .

How can eq.( 9 ) be applied to in obtaining an analytical EOS of a
degenerate Fermi gas? Inserting,at first, the known result for $\mu _0$ into
eq.( 9 ),one would get an expression showing explicitely the dependence of a
FD integral of the order 1/2 on the particle number density and
temperature.Going a step further and inserting this result into eq.(1),one
would get the required EOS,in the form of an equation relating the number
density,temperature and various known constants (such as $h$ and $k_B$
).This result has the form 
\begin{equation}
\label{(10)}%
n_e-K_1Wn_eT^{3/2}-K_2Wn_e^{-5/3}T^{11/2}-K_3Wn_e^{-3}T^{15/2}-K_4Wn_e^{-13/3}T^{19/2}=0
\end{equation}
\medskip\ where $n_e$ denotes the number density of the electrons
and all the other symbols denote different combinations of known
constants.The low temperature limit of this equation,obtained by developing
it into series in T up to terms including $T^9$ , is solvable within
S.Wolfram's MATHEMATICA 3 software package in a few minutes on a Pentium
MMX/166 with 32 Mbytes RAM. The solution has the form

\begin{equation}
\label{(11)}n_{}\cong -\frac 1{6^{1/4}}[[\frac{%
6K_3WT^{15/2}-12K_1K_3W^2T^9+6K_1^2K_3W^3T^{21/2}}{%
1-3K_1WT^{3/2}+3K_1^2T^3-K_1^3W^3T^{9/2}}-<<...>> 
\end{equation}
\medskip\ where $<<..>>$ denotes terms
omitted due to space limitations.

We have thus obtained an analytical form of the EOS of a degenerate Fermi
gas.The volume per particle,needed in some applications,is simply the
inverse density.Our result is approximate,in the sense that the developement
in eq.( 9 ) is limited to a small number of terms.Increasing the number of
terms and applying this EOS to real physical systems will be the subject of
future work.\medskip\ 

\begin{center}
Acknowledgement
\end{center}

I am grateful to Dr.Gradimir Vujicic from the Vinca Institute of Nuclear
Sciences for a discussion which motivated this work,and to Mr. Kristijan
Lazic for the installation of the computing facility.\medskip\ 

References

Antia,H.M.: Astrophys.J.Suppl.Ser.,{\bf 84},101 ( 1993 ).

Chandrasekhar,S.: Stellar Structure,Univ.of Chicago Press.,Chicago

( 1939 ).

Cloutman,L.D.: Astrophys.J.Suppl.Ser.,{\bf 71}, 677 ( 1989 ).

Landau,L.D.and Lifchitz,E.M.: Statistical Physics,Vol.1,Nauka,Moscow

( 1976 ).

Miralles,J.A.and Van Riper,K.A.: preprint astro-ph/9509124 ( 1995 ).

\end{document}